\documentclass[preprint,12pt]{elsarticle}

\usepackage{subcaption}
\usepackage{lineno,hyperref}
\usepackage{natbib}
\usepackage{geometry}
\usepackage{fleqn}
\usepackage{graphicx}
\usepackage{newtxtext,newtxmath}
\usepackage{hyperref}
\usepackage{booktabs}
\usepackage{bm}
\usepackage{amsmath}
\usepackage{enumitem}
\usepackage{threeparttable}
\usepackage{esvect}
\usepackage{soul}
\usepackage{upgreek}
\modulolinenumbers[5]

\journal{Elsevier}

\usepackage{caption}
\usepackage[table]{xcolor}
\usepackage{siunitx}

\usepackage{etoolbox}
\appto\TPTnoteSettings{\footnotesize}

\begin{document}

\begin{frontmatter}

\title{\large{Deployment of Traditional and Hybrid Machine Learning for Critical Heat Flux Prediction in the CTF Thermal Hydraulics Code}}

\author[NCSU]{Aidan Furlong\corref{mycorrespondingauthor}}
\cortext[mycorrespondingauthor]{Corresponding author}
\ead{ajfurlon@ncsu.edu}

\author[UTK]{Xingang Zhao}

\author[ORNL]{Robert K. Salko}

\author[NCSU]{Xu Wu}

\address[NCSU]{Department of Nuclear Engineering, North Carolina State University    \\ 
	Burlington Engineering Laboratories, 2500 Stinson Dr., Raleigh, NC 27695 \\}

\address[UTK]{Department of Nuclear Engineering, University of Tennessee, Knoxville, \\ Zeanah Engineering Complex, 863 Neyland Dr., Knoxville, TN 37916}

\address[ORNL]{Nuclear Energy and Fuel Cycle Division, Oak Ridge National Laboratory, \\ 1 Bethel Valley Rd., Oak Ridge, TN, USA 37830}

\begin{abstract}
Critical heat flux (CHF) marks the transition from nucleate to film boiling, where heat transfer to the working fluid can rapidly deteriorate. Accurate CHF prediction is essential for efficiency, safety, and preventing equipment damage, particularly in nuclear reactors. Although widely used, empirical correlations frequently exhibit discrepancies in comparison with experimental data, limiting their reliability in diverse operational conditions. Traditional machine learning (ML) approaches have demonstrated the potential for CHF prediction but have often suffered from limited interpretability, data scarcity, and insufficient knowledge of physical principles. Hybrid model approaches, which combine data-driven ML with physics-based models, mitigate these concerns by incorporating prior knowledge of the domain. This study integrated a purely data-driven ML model and two hybrid models (using the Biasi and Bowring CHF correlations) within the CTF subchannel code via a custom Fortran framework. Performance was evaluated using two validation cases: a subset of the Nuclear Regulatory Commission CHF database and the Bennett dryout experiments. In both cases, the hybrid models exhibited significantly lower error metrics in comparison with conventional empirical correlations. The pure ML model remained competitive with the hybrid models. Trend analysis of error parity indicates that ML-based models reduce the tendency for CHF overprediction, improving overall accuracy. These results demonstrate that ML-based CHF models can be effectively integrated into subchannel codes and can potentially increase performance in comparison with conventional methods.
\end{abstract}

\begin{keyword}
critical heat flux, hybrid modeling, machine learning
\end{keyword}

\end{frontmatter}

{\renewcommand\thefootnote{}\footnotetext{This manuscript has been authored by UT-Battelle LLC, under contract DE-AC05-00OR22725 with the US Department of Energy (DOE). The US government retains and the publisher, by accepting the article for publication, acknowledges that the US government retains a nonexclusive, paid-up, irrevocable, worldwide license to publish or reproduce the published form of this manuscript, or allow others to do so, for US government purposes. DOE will provide public access to these results of federally sponsored research in accordance with the DOE Public Access Plan (http://energy.gov/downloads/doe-public-access-plan).}}

\section{Introduction}

Critical heat flux (CHF) marks the transition from nucleate to film boiling, where the walls of a heated channel become coated with vapor instead of liquid coolant. This phase change significantly reduces heat transfer, causing a rapid increase in surface temperature due to the insulating effect of the vapor layer. Such a transition can deteriorate system efficiency and lead to equipment damage. CHF is a key consideration in applications such as steam generators, microelectronics cooling, and chemical processing. In nuclear systems, CHF is a critical safety parameter for preventing fuel bundle failure and protecting the integrity of ceramic fuel.

Over the past century, research on CHF has primarily involved experiments in heated tubes under well-controlled thermal hydraulic conditions. This long history has produced hundreds of empirical correlations, such as the Biasi, Bowring, CISE-4, and W-3 correlations \cite{todreas2021nuclear}, each tailored to specific CHF mechanisms or flow conditions. An alternative to correlations is the 2006 Groeneveld look-up table (LUT) \cite{groeneveld2019critical}, which was developed from nearly 25,000 measurements from experiments in vertically oriented, uniformly heated tubes across a wide range of conditions. Although extensively used in the nuclear industry, both the Groeneveld LUT and the empirical correlations exhibit discrepancies in certain operational regimes \cite{groeneveld20072006}, motivating ongoing efforts to develop more precise CHF prediction methods.

In recent years, machine learning (ML) techniques---including deep neural networks (DNNs), convolutional neural networks (CNNs), support vector machines (SVMs), and random forests (RFs)---have been introduced as promising alternatives for CHF prediction \cite{jiang2013combination}\cite{kim2021prediction}\cite{zubair2022critical}\cite{helmryd2024investigation}\cite{YANG2024123167}\cite{qi2025machine}. These data-driven models can learn complex relationships within high-dimensional datasets that conventional empirical correlation based approaches may omit. Once training is complete, the models are lightweight and computationally efficient when making predictions. Recognizing these benefits, the Organisation for Economic Co-operation and Development Nuclear Energy Agency (OECD NEA) established the Task Force on Artificial Intelligence and Machine Learning for Scientific Computing in Nuclear Engineering in 2022 with an initial focus on benchmarking ML-based CHF prediction using input parameters such as diameter ($D$), heated length ($L$), pressure ($P$), mass flux ($G$), and outlet equilibrium quality ($x_\text{e}$) \cite{nea2024benchmark}.

Despite their advantages, ML models often suffer from issues with interpretability. Interpretability is a critical concern in nuclear engineering because reliability and transparency are requirements. These models typically contain thousands or even millions of parameters, which can obscure their decision-making processes to developers. To mitigate this challenge, a hybrid approach was proposed \cite{zhao2020prediction} that integrates a well-established thermal hydraulics correlation as a ``base model'' coupled with an ML model to correct the outputs to better match experimental observations. In hybrid approaches such as this, much of the physical knowledge of CHF is provided by the base model, which reduces the amount of inferred knowledge required in the ML component. Classified as a parallel ``gray-box'' method \cite{thompson1994modeling}, these hybrid approaches increase the model's interpretability, stability, and resistance to performance degradation caused by limited training data \cite{psichogios1992hybrid}.

Recent studies have further evaluated these hybrid strategies. One comparison investigated RF, DNN, and CNN models that employed the Biasi, Groeneveld LUT, or Zuber correlations, using inputs such as $D$, $L$, $P$, $G$, $x_\text{e}$, inlet temperature ($T_{\text{inlet}}$), and inlet subcooling ($\Delta h_{\text{sub}}$), along with a sensitivity analysis for aleatoric uncertainty quantification \cite{mao2024uncertainty}. Similarly, another study demonstrated improved prediction accuracy when hybrid models based on artificial neural networks, SVMs, or RFs were paired with the Groeneveld LUT as a base model ; in this study, hybrid models outperformed both the standalone base model and the purely data-driven models in every case \cite{khalid2023comparison}.

A continuation study \cite{furlong2024hybrid} to Zhao et al.~\cite{zhao2020prediction} applied a hybrid approach, incorporating the Biasi and Bowring CHF empirical correlations \cite{todreas2021nuclear} as base models and evaluating their performance against a purely data-driven DNN across various training set sizes. In all eight cases, the hybrid models outperformed the pure DNN, and the greatest performance gap was observed in the most data-restrictive scenario. When the training set was limited to just nine points, the pure DNN’s relative error increased to 48.25\%. Conversely, both hybrid models maintained superior accuracy, achieving relative errors below 6.40\% (slightly smaller than the standalone empirical correlations).

Although both purely data-driven and hybrid CHF models are well-studied in a research setting, little work has been conducted to test their performance in practice. In the field of nuclear engineering, several thermal hydraulics codes exist to facilitate the design of nuclear reactor cores and to provide safety analysis \cite{trace_theory}\cite{retran}\cite{thurgoodcobra}\cite{salko2024ctf}. Many of these codes frequently use empirical CHF models, such as the Biasi and Bowring correlations. The authors are currently unaware of any pre-existing literature that describes the deployment of CHF ML models within a thermal hydraulics code.

This work successfully demonstrates a native implementation of both purely data-driven and hybrid ML models in the CTF subchannel thermal hydraulics code. Independent test series were used to validate the efficacy of the ML models in comparison with the standalone Biasi and Bowring correlations. This paper is organized as follows: Section \ref{sec:background} summarizes key background information on CHF and relevant empirical correlations; Section \ref{sec:methods} details the hybrid ML framework, data processing steps, and the three ML models used; Section \ref{sec:results} presents results for each of the ML models when implemented in CTF, followed by a direct comparison of their performance in the two validation cases; and Section \ref{sec:conclusions} provides concluding remarks and potential future directions.

\section{Background}
\label{sec:background}

\subsection{Critical Heat Flux}
\label{subsec:critical_heat_flux}

The study of what is now known as CHF extends back as far as 1888. This phenomenon was first quantified in marine evaporators by Charles Lang \cite{lang1888temperature}. Lang's tube-shell experiments observed that continually increasing the tube surface temperature beyond a certain point led to a reduction in the ``heat units transmitted per square foot of heating area.'' Although the specific mechanism behind this observation was not identified until later, its deleterious effects on boiler efficiency motivated continued work. The current term, \textit{critical heat flux}, was introduced 71 years later by Novak Zuber in his 1959 study of nucleate and transition boiling \cite{zuber1959hydrodynamic}. This term has become common in describing the onset of the boiling crisis phenomenon.

It is now understood that CHF can present in two major phenomena: departure from nucleate boiling (DNB) and dryout (DO). The former occurs in low-quality regimes, such as those in pressurized water reactors, and is driven by the formation of a vapor film that coats the channel walls. This insulating layer significantly decreases the flow of heat from the surface to the liquid coolant, increasing the wall temperature. In higher-quality flows, such as those in boiling water reactors, a different event occurs. Instead of a predominantly liquid-filled channel, an annular flow regime develops that is characterized by a liquid annulus, gas core, and entrained droplets. The surrounding film adheres to the channel walls, providing continued cooling until the film thickness approaches zero, at which point DO occurs. Both DNB and DO can lead to unstable, high wall temperatures. In the case of Charles Lang's evaporators, the CHF phenomenon caused a reduction in efficiency, but in a high-consequence application (e.g., nuclear fuel bundles), it can cause fuel cladding failure and safety limit violation.

To predict CHF, researchers have long relied on a combination of thermal hydraulics fundamentals and physical experiments to develop empirical correlations. Hundreds of these correlations have been created for various geometries, operating conditions, working fluids, and other aspects of CHF occurrence. Two correlations that are frequently used across thermal hydraulics codes are the Biasi \cite{biasi1967studies} and Bowring \cite{bowring1972simple} correlations. The Biasi model is included in COBRA/TRAC \cite{thurgoodcobra}, TRACE \cite{trace_theory}, BISON \cite{hales2016bison}, and CTF \cite{salko2024ctf}, among others. This correlation is capable of predicting CHF in cases of both DNB and DO, performing with a root mean square error of 7.26\% on its 4,500 experimental data points \cite{todreas2021nuclear}. The Bowring correlation is used in COBRA/TRAC and CTF and is valid for a wide range of mass fluxes and pressures, performing with a root mean square error of 7\% on its 3,800 data points. For the purposes of this study, the Biasi and Bowring correlations were both used as base models for the hybrid approaches and as a reference solution for performance comparison.


\subsection{Nuclear Regulatory Commission CHF Database}
\label{subsec:nrc_database}

The dataset used to construct the 2006 Groeneveld LUT \cite{groeneveld2019critical}, hereinafter referred to as the public Nuclear Regulatory Commission (NRC) CHF database, contains 24,579 points from physical experiments performed over the course of 60 years. All these experiments used water-cooled cylindrical tubes, with a large range of CHF values ranging from 50 to 16,339 \si{\kilo\watt\per\square\meter} over a diverse set of operational conditions. The NRC CHF database contains both measured and derived parameters: tube diameter ($D$), heated length ($L$), pressure ($P$), mass flux ($G$), outlet equilibrium quality ($x_\text{e}$), inlet subcooling ($\Delta h_{\text{in}}$), and inlet temperature ($T_{\text{in}}$). As the largest publicly available compilation of CHF data, this database is a practical choice for building data-driven ML models. As further discussed in Section \ref{subsec:data_preparation}, most of these data were used for the training and optimization of this study's ML-based models, and a portion was also preserved for the evaluation of the final models. Because these ``test'' data are never seen during training, they are an appropriate measure of how well a trained model generalizes to new data.

\subsection{Bennett Experiments}

The Bennett test series was a set of experiments that considered DO. The work, originally published in 1967 by the United Kingdom Atomic Energy Authority \cite{bennett1967paper}, was performed at the Harwell High Pressure Two-Phase Heat Transfer Loop facility. These tests focused on the vertical upflow of water within heated tubes of a constant diameter (12.62 \si{\milli\meter}) made from joined sections of Nimonic Alloy 80 tubing. Two heated lengths were considered (3.66 \si{\meter} and 5.56 \si{\meter}), both with a uniform heat flux profile and a series of 27 thermocouples running axially. Each of the 224 experimental cases was performed with a stable mass flux in the range of 393 to 5,235 \si{\kilo\gram\per\square\meter\per\second} and inlet subcooling in the range of 42 to 181 \si{\kilo\joule\per\kilo\gram}. The uncertainty associated with inlet subcooling was noted to be $\pm 23.26$ \si{\kilo\joule\per\kilo\gram}.

Operationally, system power was gradually increased until a temperature excursion on the exit thermocouple was observed, indicating DO. The system was then returned to a steady state, and then power was again raised until the next thermocouple reported a sharp rise in temperature. This process continued until any one of the thermocouples reached a cutoff setpoint of 1,400 \si{\degree F} for equipment protection purposes. The included dataset reports temperature values for each of the 27 thermocouples.

This test series was not included in the public NRC CHF database, allowing it to be used independently as a blind test set for evaluating the ML models. Compared with the previous dataset of Section \ref{subsec:nrc_database}, which consists of points from many physical experiments and different facilities, the Bennett dataset can help validate the ML methods in the lens of a single test series with similar conditions.

Table \ref{tab:test_ranges} summarizes the ranges of experimental conditions between the training data and the two test series.  The original units from Bennett were converted to those used in the NRC CHF database as appropriate; relevant fluid properties were computed using the IAPWS-IF67 formulation. Although a simple min--max inspection suggests that both the NRC test dataset and the Bennett test series lie within the training data’s interpolating regime, this approach does not account for feature correlations. To rigorously confirm interpolation, principal component analysis (PCA) and convex hull analysis were applied. PCA transforms the feature space into an orthogonal coordinate system where each principal component is a linear combination of the original features ordered by variance. Dimensionality reduction can be achieved by selecting only the top components, which capture most of the variance. Convex hull analysis then identifies the smallest enclosing boundary around the training data in this transformed space, defining the interpolating region.

\renewcommand{\arraystretch}{1.2}
\begin{table}[htb!]
    \centering
    \begin{threeparttable}
    \caption{Parameter ranges for this study's training data and two test cases. Some parameters are constant throughout a study and are listed twice under ``Min'' and ``Max''.}
    \label{tab:test_ranges}
    \begin{tabular}{l cc cc cc}
        \toprule
        Parameter & \multicolumn{2}{c}{\textbf{NRC (Train)}} & \multicolumn{2}{c}{\textbf{NRC (Test)}} & \multicolumn{2}{c}{\textbf{Bennett}} \\
        & Min & Max & Min & Max & Min & Max \\
        \midrule
        $D$ (\si{\milli\meter}) & $2.00$ & $16.00$ & $2.00$ & $16.00$ & $12.62$ & $12.62$ \\ \hline
        $L$ (\si{\meter}) & $0.05$ & $20.00$ & $0.05$ & $20.00$ & $3.66$ & $5.56$ \\ \hline
        $P$ (\si{\kilo\pascal}) & $100$ & $20{,}000$ & $100$ & $20{,}000$ & $6{,}895$ & $6{,}895$\\ \hline
        $G$ (\si{\kilo\gram\per\square\meter\per\second}) & $8$ & $7{,}964$ & $10.6$ & $7{,}961$ & $393$ & $5{,}235$ \\ \hline
        $x_\text{e}$ (--) & $-0.50$ & $0.99$ & $-0.46$ & $0.99$ & $0.17$\tnote{a} & $0.88$\tnote{a} \\ \hline
        $\Delta h_{\text{sub}}$ (\si{\kilo\joule\per\kilo\gram}) & $-1{,}211$ & $1{,}644$ & $-1{,}178$ & $1{,}549$ & $42$ & $181$ \\ \hline
        $T_{\text{in}}$ (\si{\celsius}) & $5.89$ & $361.94$ & $8.50$ & $356.49$ & $284$\tnote{a} & $284$\tnote{a} \\ \hline
        $q''_{\text{cr}}$ (\si{\kilo\watt\per\square\meter}) & $50$ & $16{,}339$ & $90$ & $13{,}850$ & $362$ & $2{,}074$ \\
        \bottomrule
    \end{tabular}
    \begin{tablenotes}
    \item[a] Values derived from other parameters and IAPWS-IF97 fluid properties.
    \end{tablenotes}
    \end{threeparttable}
\end{table}

\begin{figure}[htb!]
    \centering
    \begin{subfigure}{0.49\textwidth}
        \centering
        \includegraphics[width=\linewidth]{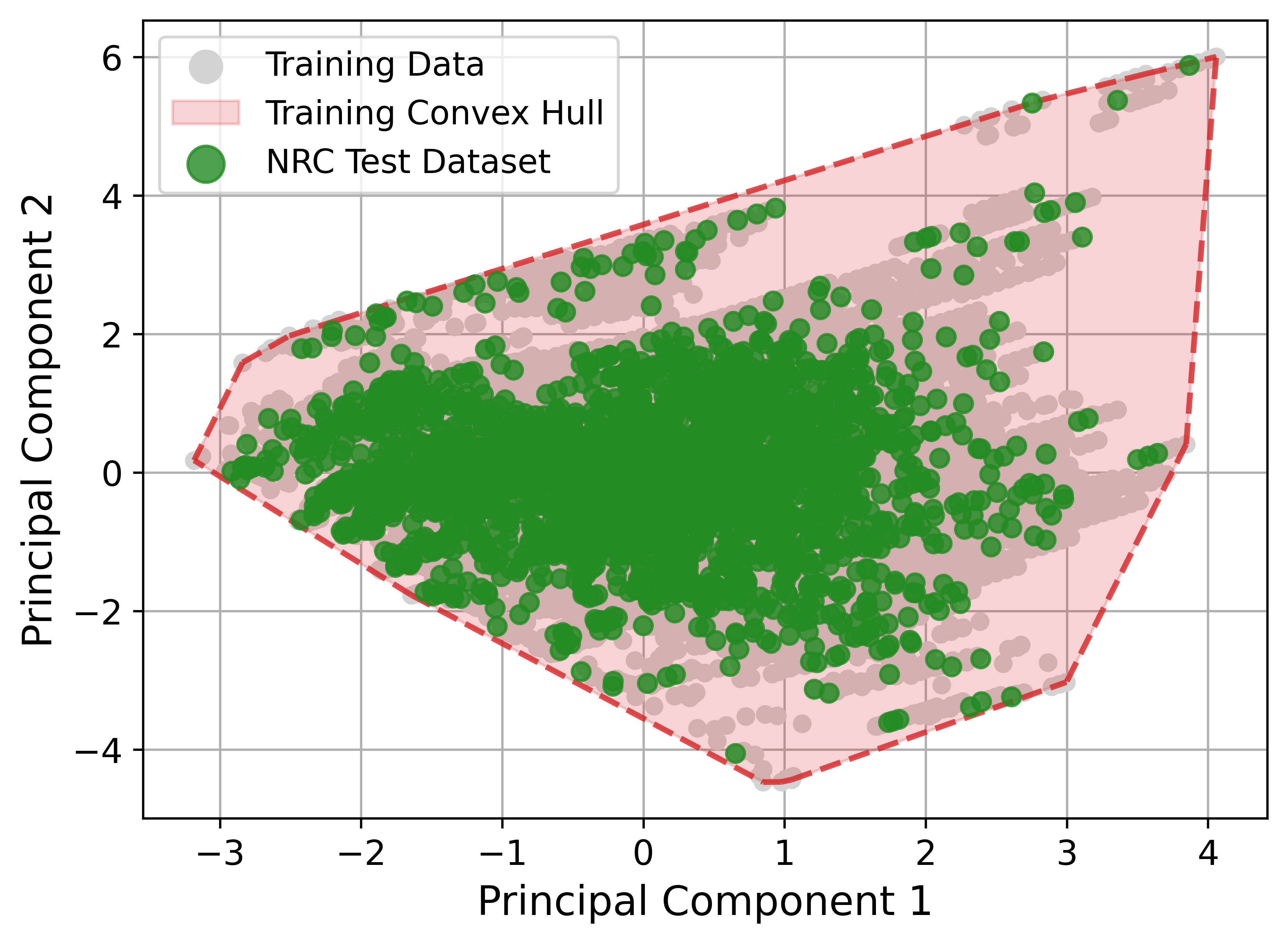}
        \caption{NRC Test Dataset}
        \label{subfig:nrc_pca_convex}
    \end{subfigure}
    \begin{subfigure}{0.49\textwidth}
        \centering
        \includegraphics[width=\linewidth]{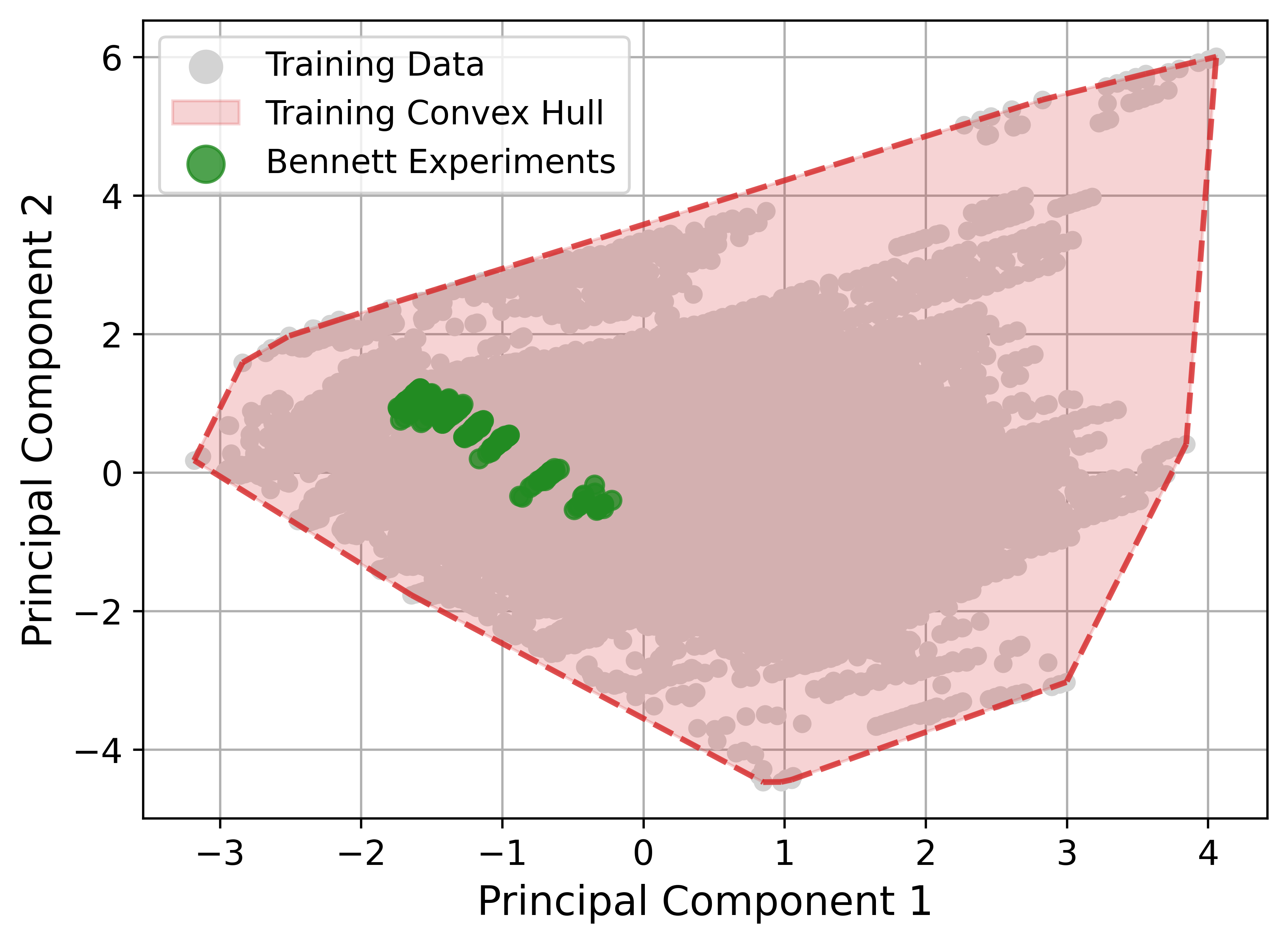}
        \caption{Bennett Experiments}
        \label{subfig:bennett_pca_convex}
    \end{subfigure}
    \caption{Two-component PCA paired with convex hull analysis visualizing the evaluation data in an interpolating regime. The NRC test dataset's distribution is similar to that of the training data, with coverage extending to the extrema. The Bennett test series are well enclosed by the boundary and are located in a region of dense training data.}
    \label{fig:pca_convex}
\end{figure}

A point is classified as interpolated if it lies within the convex hull and is classified as extrapolated otherwise. This distinction is important for ML models because performance tends to degrade in extrapolated domains. Although two principal components are used for visualization in Figure \ref{fig:pca_convex}, convex hull analysis was also performed in the original feature space to ensure accuracy. The results show that all points in the Bennett test series fall within the interpolating regime, whereas 66 points (of 2,458) from the NRC testing dataset lie outside the convex hull, indicating extrapolation.

\section{Methods}
\label{sec:methods}

\begin{figure}[htb!]
    \centering
    \includegraphics[width=0.9\linewidth]{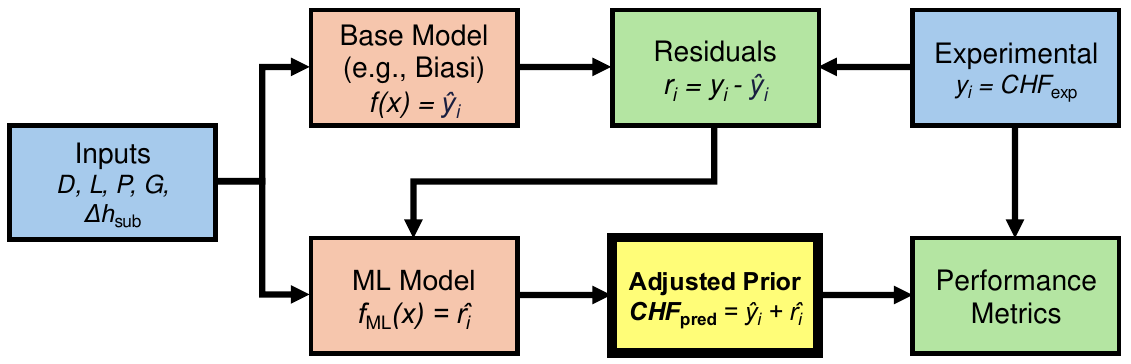}
    \caption{Workflow for the hybrid model methodology in the \textit{training} configuration. Instead of directly predicting CHF values, the ML model component predicts the difference (residual) between the base model's CHF output and the experimental CHF value. After training, in the \textit{deployment} configuration, actual residual values are unknown, so it is entirely up to ML model to adjust the base model's output only with knowledge of the corresponding input vector.}
    \label{fig:flowchart_hybrid}
\end{figure}

The fundamental concept behind the proposed knowledge-informed hybrid ML approach involves combining two models: a physics-based base model that provides an initial estimate and a data-driven ML model that refines this estimate. In this study, the base model is represented by either the Biasi or Bowring empirical CHF correlations, which yield a predicted CHF value ($\hat{y}$) that may deviate from experimentally measured values ($y$). Because the correlations are solved using the heat balance method (HBM) described in Section \ref{subsec:data_preparation}, the chosen input parameters include $D$, $L$, $P$, $G$, and $\Delta h_{\text{sub}}$. After computing the CHF value with the correlation, the ML model estimates the expected residual ($\hat{r}$) between this value and the corresponding experimental measurement. Incorporating the ML-predicted residual into the initial CHF estimate provides the final corrected CHF prediction ($\widetilde{y}$). Figure \ref{fig:flowchart_hybrid} illustrates this workflow in its training configuration, where experimental CHF values serve as the ground truth, enabling the ML model to learn the true residual ($r$) as its target.

\subsection{Data Preparation} \label{subsec:data_preparation}

The full 24,579-entry public NRC CHF database was used for training, validation, and testing of the pure and hybrid ML models. The first step in data pre-processing concerned the generation of the base model outputs as well as their residuals against the experimental CHF values. Both the Biasi and Bowring correlations are functions of $D$, $G$, $P$, and critical equilibrium quality ($x_{\text{cr}}$), having a particular sensitivity to the latter. The NRC database's outlet $x_\text{e}$ values are equivalent to $x_{\text{cr}}$, as CHF occurs at the tube's outlet for these experiments. There are two established methods for solving these correlations for CHF: the direct substitution method (DSM) and the HBM \cite{hejzlar1996consideration}. The DSM, which is also known as the ``constant local condition'' method, simply uses the local quality directly, whereas the HBM uses constant inlet conditions and a heat balance, taking into account channel height. The HBM iteratively solves the heat balance and correlation, varying heat flux until an exact critical condition is found at a specified height. There has been discourse regarding which method is more appropriate in certain situations, and most studies indicate that the HBM should be used whenever possible \cite{filho2003margin}. One limitation that restricts the generality of the HBM is its inaccuracy in cases of quick transients; in these cases, comparing both HBM and DSM values has been recommended. In cases of steady-state and quasi-steady-state conditions, the HBM shows superior agreement with experimental results in comparison with the DSM \cite{celata1996application}. Notably, the original formulation of the Bowring correlation \cite{bowring1972simple} uses the constant inlet condition approach (but without iteration) and yields nearly identical results in comparison with the explicit HBM implementation.

In this study, instead of using the quality provided within the database, $x_{\text{cr}}$ was computed simultaneously \cite{todreas2021nuclear} with CHF. This approach uses the correlation equation (either Biasi or Bowring) and the heat balance provided in Eq.~(\ref{eqn:heat_balance}). The HBM in this case requires three additional parameters in comparison with the DSM: the critical length from the inlet ($L_{\text{cr}}$), $\Delta h_{\text{sub}}$, and the latent heat of vaporization ($h_{\text{fg}}$). The first two are provided in the NRC database ($L$ being equal to $L_\text{cr}$), and $h_{\text{fg}}$ can be computed based on system pressure using fluid properties from the IAPWS-IF97 formulation.

\begin{equation}\label{eqn:heat_balance}
    x_{\text{cr}} = x(L_{\text{cr}}) = \frac{q_{\text{cr}}}{\dot{m} h_{\text{fg}}} - \frac{\Delta h_{\text{sub}}}{h_{\text{fg}}} = \frac{4 q''_{\text{cr}} L_{\text{cr}}}{G D h_{\text{fg}}} - \frac{\Delta h_{\text{sub}}}{h_{\text{fg}}}
\end{equation}

Once the correlation outputs and residuals were added to the working dataset, it was then shuffled to promote homogenized distribution of experimental conditions and to prevent biasing of the model simply due to the order of entries. All data were then standardized (also referred to as z-score normalization) to force a distribution with a zero mean and standard deviation of one \cite{garcia2015data}. This was done to reduce the likelihood of biasing a model due to differences in the absolute magnitudes of input features (such as tube diameter in comparison with the much larger mass flux). The data were then split into training, validation, and testing partitions using an 80\%/10\%/10\% ratio. The validation partition was used throughout the optimization process of model architecture and hyperparameter configuration and was not used during final model evaluation. If the holdout testing set were to be used during the optimization process, knowledge of its values \textit{could} be transferred through the hyperparameters themselves, potentially inflating performance metrics \cite{boulesteix2015ten}.

\subsection{ML Model Structure} \label{subsec:architecture}

Each of the three ML models was of identical architecture: seven hidden layers (of widths 44, 64, 41, 26, 67, 10, and 17 neurons) with one single-neuron output layer. The mean square error loss function was chosen in conjunction with the Adam optimizer. These models were implemented using Google's TensorFlow \cite{tensorflow2015-whitepaper} and were trained with an exponential learning rate decay to promote final model stability. A total of 16 hyperparameters (activation, neurons, learning rate, and batch size) were optimized, in addition to the model depth, using a random search involving 1,000 prospective configurations. The ranges used to construct the search space are reported in Table \ref{tab:dnn_hp_tuning}. The tuning process used RayTune \cite{liaw2018tune} and was further optimized by the asynchronous successive halving algorithm (ASHA), which accelerates tuning via the early termination of underperforming models \cite{li2020system}. ASHA evaluates using an internal validation set dynamically split from the training data, distinct from the validation partition used after hyperparameter optimization is complete. Once the best-performing hyperparameter configuration was found, the final model was trained to 500 epochs and evaluated using the shuffled test set.

\renewcommand{\arraystretch}{1.1}
\begin{table}[ht!]
\normalsize
\captionsetup{justification=centering}
\centering \caption{DNN hyperparameter search space.}
\label{tab:dnn_hp_tuning}
\begin{tabular}{l l}
\toprule

Hyperparameter & Range/Values \\ 
\midrule
Hidden Layers & $4, 5, 6, 7, 8$ \\
\hline
Neurons & $[10, 70]$  \\ 
\hline
Batch size & $8, 16, 32, 64$ \\ 
\hline
Learning rate & $[0.0001,0.01]$ \\
\hline
Epochs & $[200, 500]$ \\ 
\hline
Activation & \texttt{elu}, \texttt{relu}, \texttt{softplus}, \texttt{sigmoid}, \texttt{tanh}\\ 
\bottomrule
\end{tabular}
\end{table}

\subsection{CTF Implementation} \label{subsec:implementation}

CTF \cite{salko2024ctf} is a thermal hydraulics code developed specifically for modeling transient and steady-state behavior in nuclear reactor cores with rod-bundle geometries.  
CTF performs a numerical solution of two-phase flow using the subchannel method, which is a specialization of the porous media method that is made by assuming flow direction is dominant in one direction due to system geometry constraints.
A two-fluid, three-field approach is used for modeling two-phase flow behavior.
This results in a system of seven governing equations being solved to model two-phase flow: two mass equations (vapor and droplets), three momentum equations (vapor, droplets, and liquid film), and two energy equations (liquid and vapor).
Assuming that the volume fraction of the three fields adds to unity, a third mass equation is not required.  
Furthermore, liquid film and liquid droplets are also assumed to be in thermodynamic equilibrium, eliminating the need for a third energy equation.
In addition to the fluid solution, CTF provides a solid energy equation for modeling conduction heat transfer through solid objects, such as metal walls, and cylindrical nuclear fuel rod geometries.
CTF is developed for capturing a range of thermal hydraulic conditions possible in light water reactors during normal and off-normal behavior. 
A flow regime map is used to classify the structure of the two-phase flow (i.e., small bubbles, slug flow, annular mist), and a set of closure models is provided to determine field interactions (i.e., interfacial drag and heat transfer).
Closure models are also provided to determine heat transfer in multiple different regimes (single-phase flow, subcooled and saturated boiling, transition boiling, and film boiling).
The heat transfer regime determination is of critical importance for nuclear reactor safety modeling because of its effect on system temperatures, and this determination relies heavily on the accuracy and uncertainty of the CHF model.

Because the ML models were trained using TensorFlow, they were initially incompatible with CTF, which is primarily written in Fortran. To avoid relying on an intermediate Python helper script for per-prediction use, a native Fortran framework was developed to make predictions directly within CTF, eliminating the need for TensorFlow's C API and maintaining a consistent Fortran-based workflow \cite{furlong2024fortran}\cite{furlong2025nativefortranimplementationtensorflowtrained}. This framework extracts the model weights, biases, and relevant metadata (such as standardization parameters) from the saved TensorFlow HDF5 model files; the metadata are stored in a separate HDF5 file. Each neural network is initialized once during CTF execution, allowing subsequent predictions to be efficiently performed within the \texttt{Heatfunctions.f90} module, where the conventional CHF subroutines reside. This setup allows the ML models, both pure and hybrid, to effectively be a drop-in option for CHF prediction. Within the input deck, the user is able to simply change the CHF-solve value (\texttt{W3CHF}) to one of the three added ML-driven model options: pure ML, hybrid Biasi, or hybrid Bowring.

Both the hybrid subroutines call the standard correlations' subroutines to provide the ``base'' model estimate, which is then modified with the residual subsequently predicted by the ML model. Although this does slightly increase the time expenditure in comparison with the standalone correlations, the effect is minimal, with 1,000 ML model evaluations performed in under 8 \si{\milli\second}. The pure ML model evaluates only the ML model and avoids this additional step, as the standard CHF subroutines are not called. Another modification in CTF was to the Biasi and Bowring correlation functions to use the HBM instead of the currently implemented DSM.

\section{Results}
\label{sec:results}

Once the ML models were implemented within CTF, their Fortran-based outputs were directly verified against the TensorFlow-based predictions with a set of reference points to ensure that the framework was functioning as expected. Deviation of the Fortran implementation was minimal (on average \SI{4e-6}{\kilo\watt\per\square\meter}) and not indicative of systematic error within the module, aligning with verification results of previous cases \cite{furlong2025nativefortranimplementationtensorflowtrained}. Once this verification was complete, work proceeded with validation using the two testing datasets described in Section \ref{sec:background} to study how the ML models behave with CTF simulations instead of merely feeding them inputs directly from a database. 

To compare the performance of each method, six error metrics were chosen: mean relative error ($\upmu_{\text{error}}$), maximum relative error ($\text{Max}_{\text{error}}$), standard deviation of the error distribution ($\text{Std}_{\text{error}}$), relative root mean square error ($rRMSE$), the fraction of cases with relative error above 10\% ($F_{\text{error}} > 10\%$), and, similarly, the fraction of cases with relative error above 25\% ($F_{\text{error}} > 25\%$). The definition of \textit{rRMSE} has been inconsistent among various ML CHF articles; this study defines \textit{rRMSE} in Eq.~(\ref{eqn:rrmse}), the same as that used in the OECD NEA working paper \cite{nea2024benchmark}. Here, $N$ is the number of samples in the testing dataset, $y_i$ is the ground truth value, and $\hat{y}_i$ is the final CHF prediction, regardless of the method used.

\begin{equation} \label{eqn:rrmse}
    \text{rRMSE}\; (\%) = \sqrt{\frac{1}{N}\sum_{i = 1}^N \left(\frac{\hat{y}_i - y_i}{y_i}\right)^2} \times 100 \%
\end{equation}

\subsection{Testing Results Using the NRC CHF Database}
\label{subsec:nrc_test}

The first test case for validating the ML model implementation involved modeling the NRC CHF database's experiments in CTF. Although relatively simple, this test series was necessary for further confirmation that the ML integration within CTF functioned as expected. Because the ML models were trained on an 80\% partition of the entire dataset, with an additional 10\% being used for validation, the 10\% test partition was available for evaluation. This subset consisted of 2,458 cases, which were then represented in CTF as heated tubes with geometries and conditions identical to those in the database. The same experimental assumptions were applied: vertical upflow with uniform heating and CHF occurring at the heated length's exit.

To compare the ML models' performance against a baseline, standalone Biasi and Bowring correlations were implemented using the HBM approach. These correlations were applied to the same test cases. Each model consisted of 60 axial nodes and was run as a transient to steady state over 20 simulation seconds. Before all models were run, two simple convergence studies were performed to ensure that the spatial resolution and simulation duration were adequate to achieve steady state conditions with stable exit node values. The 2,458 models were then run. A small fraction of cases (less than 10\%) terminated early due to convergence failure; these failed cases were present in both the baseline and ML-based cases, suggesting that they were unrelated to ML substitutions. The individual optimization of each case could potentially eliminate these failed cases, but this approach was not pursued due to the abundance of passing instances. The failed cases were omitted from subsequent evaluations. Once the simulations were complete, the output datasets were collected from the final experiment states at 20 simulation seconds. Because CHF is not directly provided by CTF, two values were then extracted from the exit node of each pin: the pin surface heat flux and the departure from nucleate boiling ratio (DNBR). The pin surface DNBR is computed as the ratio of the CHF to the actual heat flux at a particular location; simply multiplying this value by the local heat flux yields the CHF. These CHF values were then evaluated against experimental measurements using the six performance metrics previously described and presented in Table \ref{tab:comparison_nrc}. A limited number of extreme outliers were observed, particularly in both Biasi cases, which significantly inflated outlier-sensitive metrics such as \textit{rRMSE}. To gain a more representative perspective of these models' performance, absolute error values above the 99.5th percentile (largest 12) were omitted when computing the mean relative error, the standard deviation of the relative error, and the \textit{rRMSE} for all five approaches. These values were \textit{not} removed for the computation of the maximum relative error or the fraction of error values above 10\%. Although the outliers of the hybrid Biasi models appear to correlate to outliers within the standalone Biasi model, further investigation is needed to mitigate their occurrence.

The results summarized in Table \ref{tab:comparison_nrc} indicate that the most immediate differences between the baseline and ML cases appear in the mean relative error values. Both hybrid models achieved significantly smaller mean errors than their base model counterparts: 5.50\% versus 20.55\% for Biasi and 2.90\% versus 8.98\% for Bowring. The pure ML model also demonstrated a low mean error value (3.31\%) in comparison with the hybrid Bowring model. The largest variability across all cases occurred in the maximum relative error, ranging from 4,630.8\% in the case of the baseline Biasi to 67.5\% in the case of the pure ML model. Both the standalone and hybrid Biasi models were observed with considerably larger values than the rest but with the hybrid counterpart's value smaller than the baseline. The hybrid Bowring model saw a maximum relative error larger than that of the standalone Bowring due to an extreme outlier at a small reference CHF value.

\begin{table}[htb!]
    \centering
    \caption{Performance metrics for the test partition of the NRC dataset. The most favorable values are shown in gray.}
    \label{tab:comparison_nrc}
    \begin{tabular}{lccccc}
        \toprule
        Metric & Base & Base & Pure & Hybrid & Hybrid \\
         & Biasi & Bowring & ML & Biasi & Bowring \\
        \midrule
        $\upmu_\text{error}$ (\%) & $20.55$ & $8.98$ & $3.31$ & $5.50$ & \cellcolor{gray!25} $2.90$ \\ \hline
        $\text{Max}_{\text{error}}$ (\%) & $4{,}629.78$ & $109.40$ & \cellcolor{gray!25} $67.51$ & $4{,}445.55$ & $606.57$ \\ \hline
        $\text{Std}_{\text{error}}$ (\%) & $85.24$ & $9.93$ & $3.73$ & $11.57$ & \cellcolor{gray!25} $3.16$ \\ \hline
        $rRMSE$ (\%) & $87.66$ & $13.39$ & $4.99$ & $12.81$ & \cellcolor{gray!25} $4.28$ \\ \hline
        $F_{\text{error}}>10\%$ (\%) & $34.17$ & $31.05$ & $4.27$ & $10.52$ & \cellcolor{gray!25}$3.19$ \\ \hline
        $F_{\text{error}}>25\%$ (\%) & $12.43$ & $7.17$ & $0.48$ & $3.49$ & $\cellcolor{gray!25} 0.27$ \\
        \bottomrule
    \end{tabular}
\end{table}

Significant improvement by the ML-based models was noted in both the \textit{rRMSE} and the fraction of error values above 10\%. The baseline Biasi model reported the largest \textit{rRMSE} value of 87.66\%, nearly 7$\times$ larger than that of the hybrid Biasi model (12.81\%). The pure ML and hybrid Bowring models were observed with values significantly smaller than both baseline models. Both the empirical correlations saw over 30\% of their points above 10\% error, significantly higher than that observed in the pure ML (4.27\%), hybrid Biasi (10.52\%), and hybrid Bowring models (3.19\%). The hybrid Bowring model achieved the best performance across most metrics, except for maximum relative error (due to an extreme outlier).

\begin{figure}[htb!]
    \centering
    \begin{subfigure}{0.49\textwidth}
        \centering
        \includegraphics[width=\linewidth]{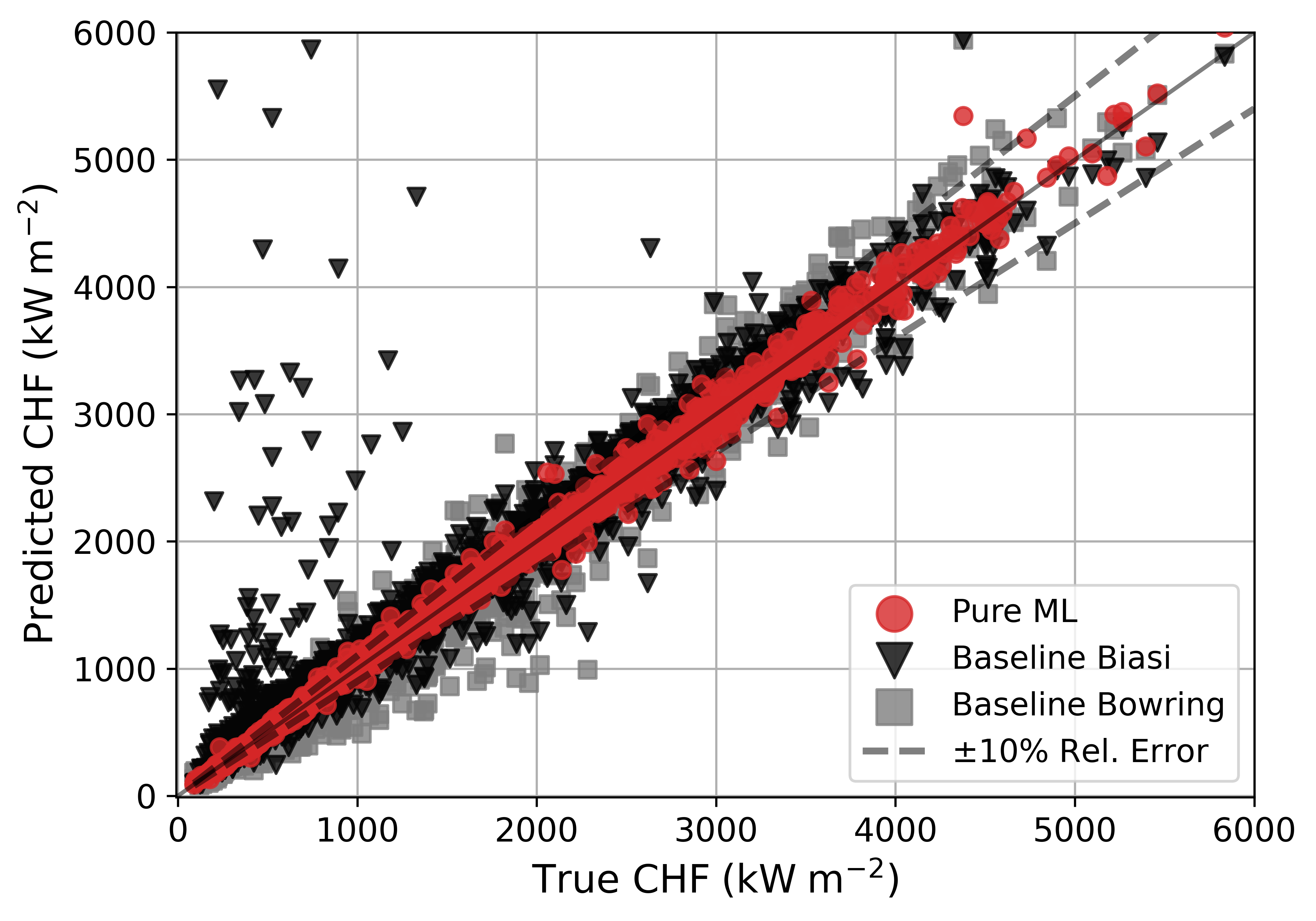}
        \caption{Parity}
        \label{subfig:parity_pure_nrc}
    \end{subfigure}
    \begin{subfigure}{0.49\textwidth}
        \centering
        \includegraphics[width=0.96\linewidth]{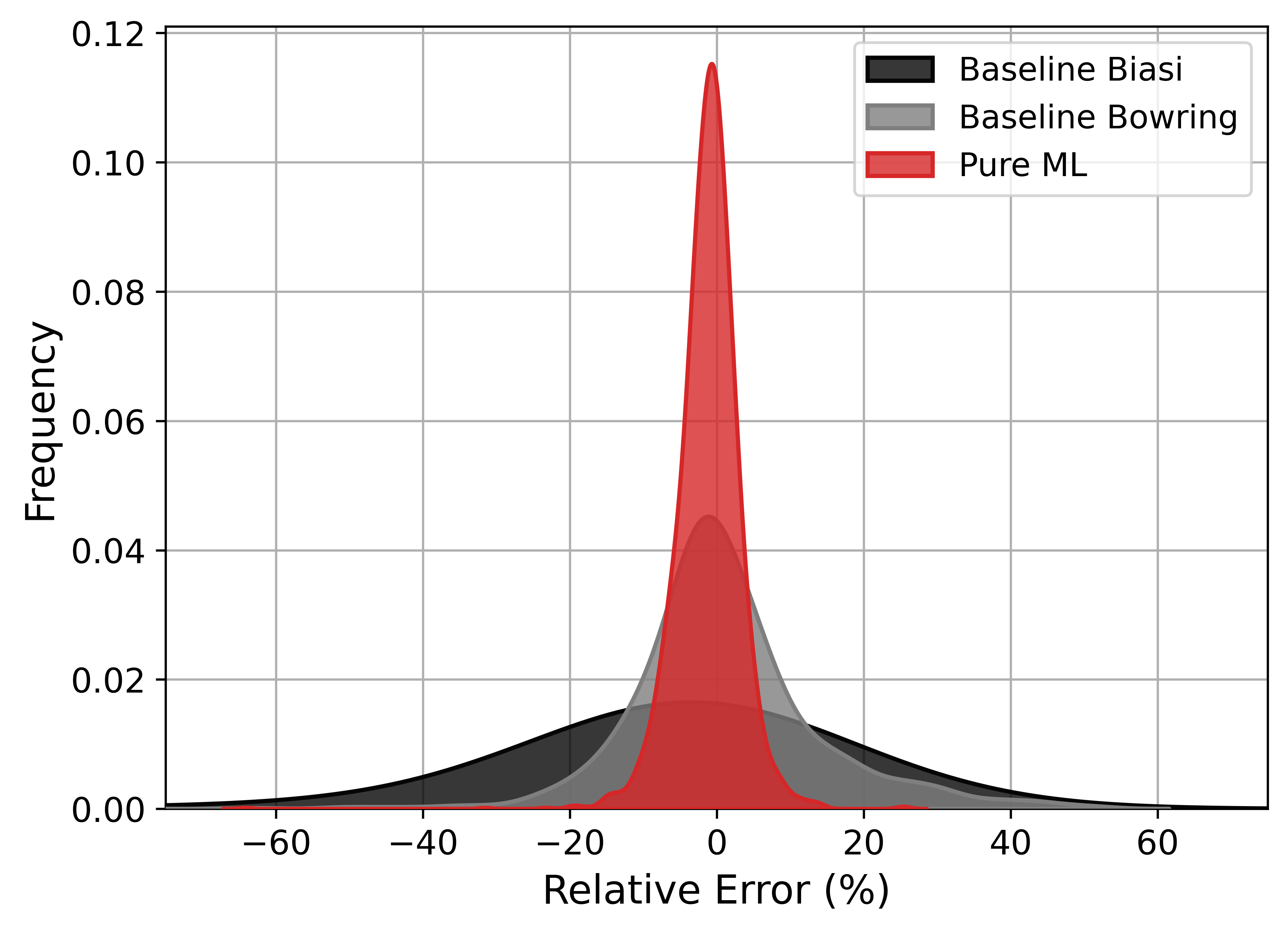} 
        \caption{KDE plot of relative error}
        \label{subfig:kde_pure_nrc}
    \end{subfigure}
    \caption{Comparisons between the baseline CHF models and the pure ML predictions using the test partition of the NRC dataset. The parity plot is windowed at 6,000 \si{\kilo\watt\per\square\meter} for ease of interpretation, and the KDE plot is windowed at $\pm 70\%$.}
    \label{fig:nrc_pure}
\end{figure}

To visualize the error distributions, the predicted and experimental CHF values were compared for parity. Figure \ref{subfig:parity_pure_nrc} shows the pure ML model's predictions against the baseline correlations. The Biasi correlation exhibited substantial overprediction at lower CHF values, producing large deviations. This behavior is attributed to cases outside the correlation’s range of validity, leading to erratic predictions. The Bowring correlation also contained cases beyond its valid range but generally produced smaller deviations, with a tendency to underpredict CHF for experimental values between 1,000 and 2,000 \si{\kilo\watt\per\square\meter}. By contrast, the pure ML model’s predictions remained centered along the identity line with a relatively small error spread.

Kernel density estimation (KDE) plots for each of the models' error distributions were then constructed for the pure and baseline cases, as shown in Figure \ref{subfig:kde_pure_nrc}. A KDE plot is used as a smoothed visualization of the relative error's probability distribution, allowing for the inspection of smoothness, spread, and central tendencies without requiring discrete histogram bins. Determination of a suitable bandwidth was completed via Silverman's rule. Due to the comparatively large maximum error values of the Biasi model, this plot was windowed at $\pm 70\%$ for ease of interpretation. The pure ML model's relative error values are shown as the tightest distribution, centered just below zero with a small tendency to overpredict CHF values. The Bowring and Biasi standalone correlations have larger spreads; the Biasi correlation has the largest spread and the greatest tendency to overpredict. Although the center of the Bowring distribution is left of zero, there is a right asymmetry corresponding to the underprediction observed below 2,000 \si{\kilo\watt\per\square\meter} in Figure \ref{subfig:parity_pure_nrc}.

\begin{figure}[htb!]
    \centering
    \begin{subfigure}{0.49\textwidth}
        \centering
        \includegraphics[width=\linewidth]{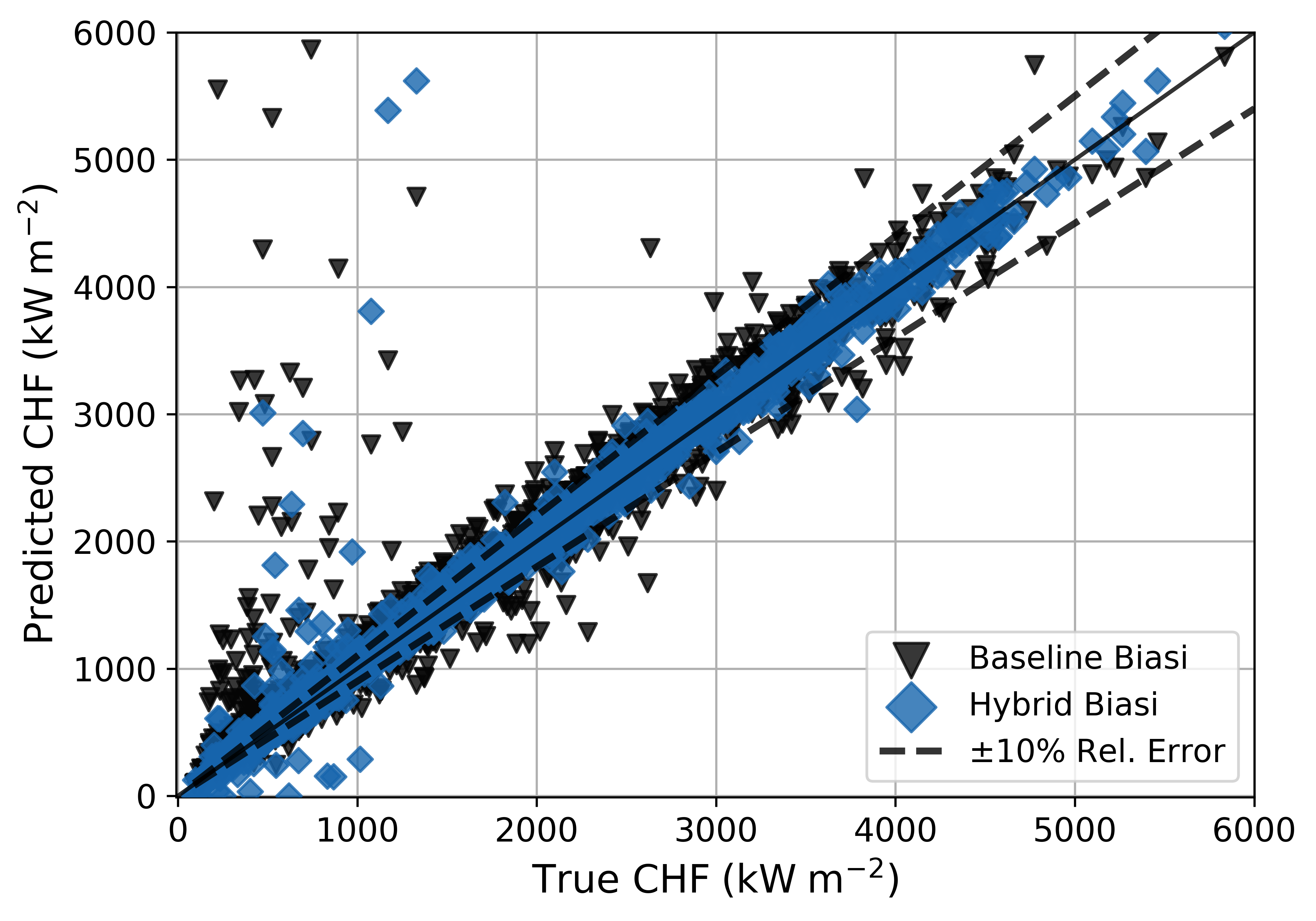}
        \caption{Parity}
    \end{subfigure}
    \begin{subfigure}{0.49\textwidth}
        \centering
        \includegraphics[width=0.96\linewidth]{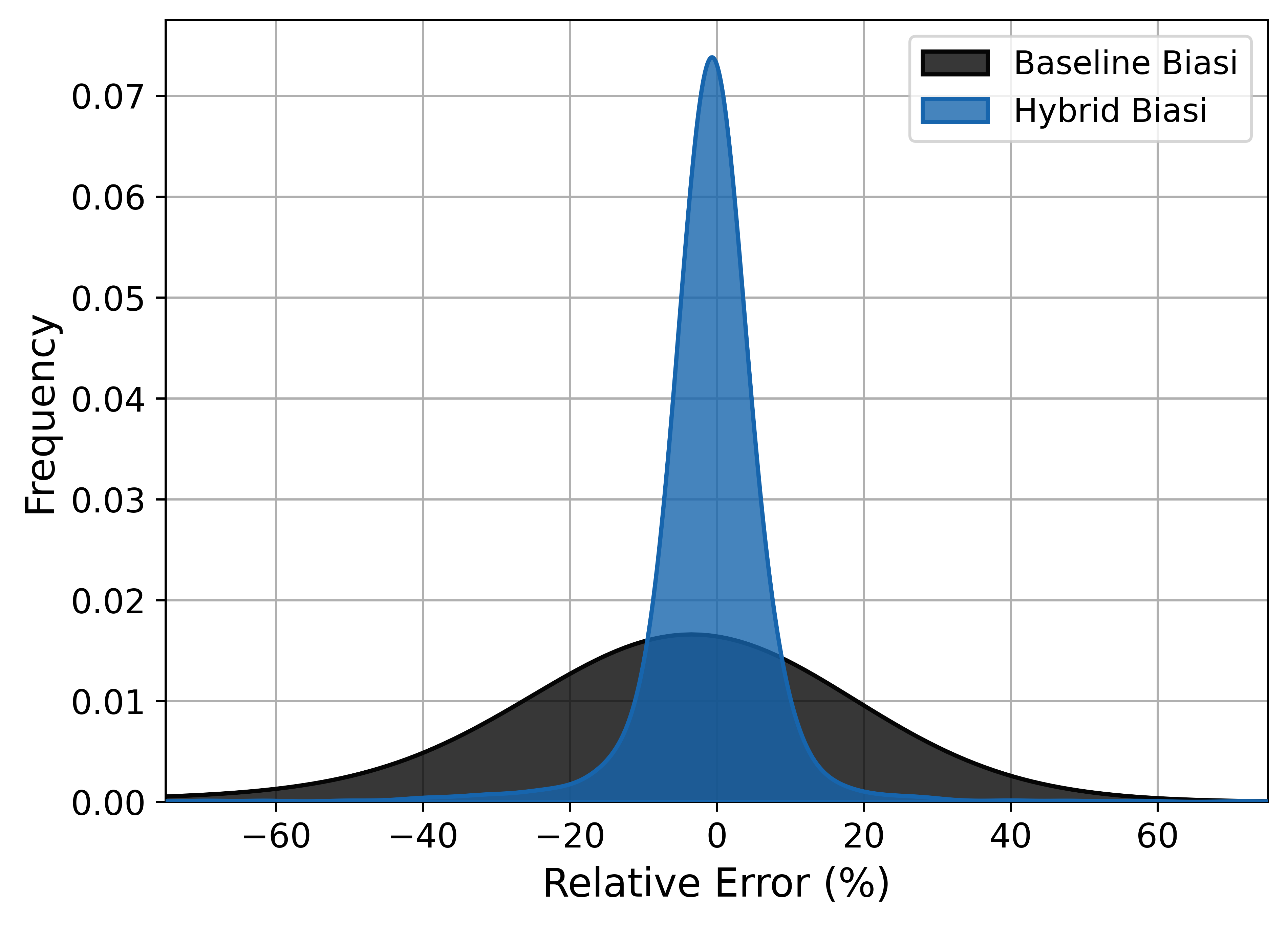}
        \caption{KDE plot of relative error}
    \end{subfigure}
    \caption{Comparisons between the baseline Biasi CHF model and its hybrid counterpart using the test partition of the NRC dataset. The parity plot is windowed at 6,000 \si{\kilo\watt\per\square\meter} for ease of interpretation, and the KDE plot is windowed at $\pm 70\%$.}
    \label{fig:nrc_biasi}
\end{figure}

\begin{figure}[htb!]
    \centering
    \begin{subfigure}{0.49\textwidth}
        \centering
        \includegraphics[width=\linewidth]{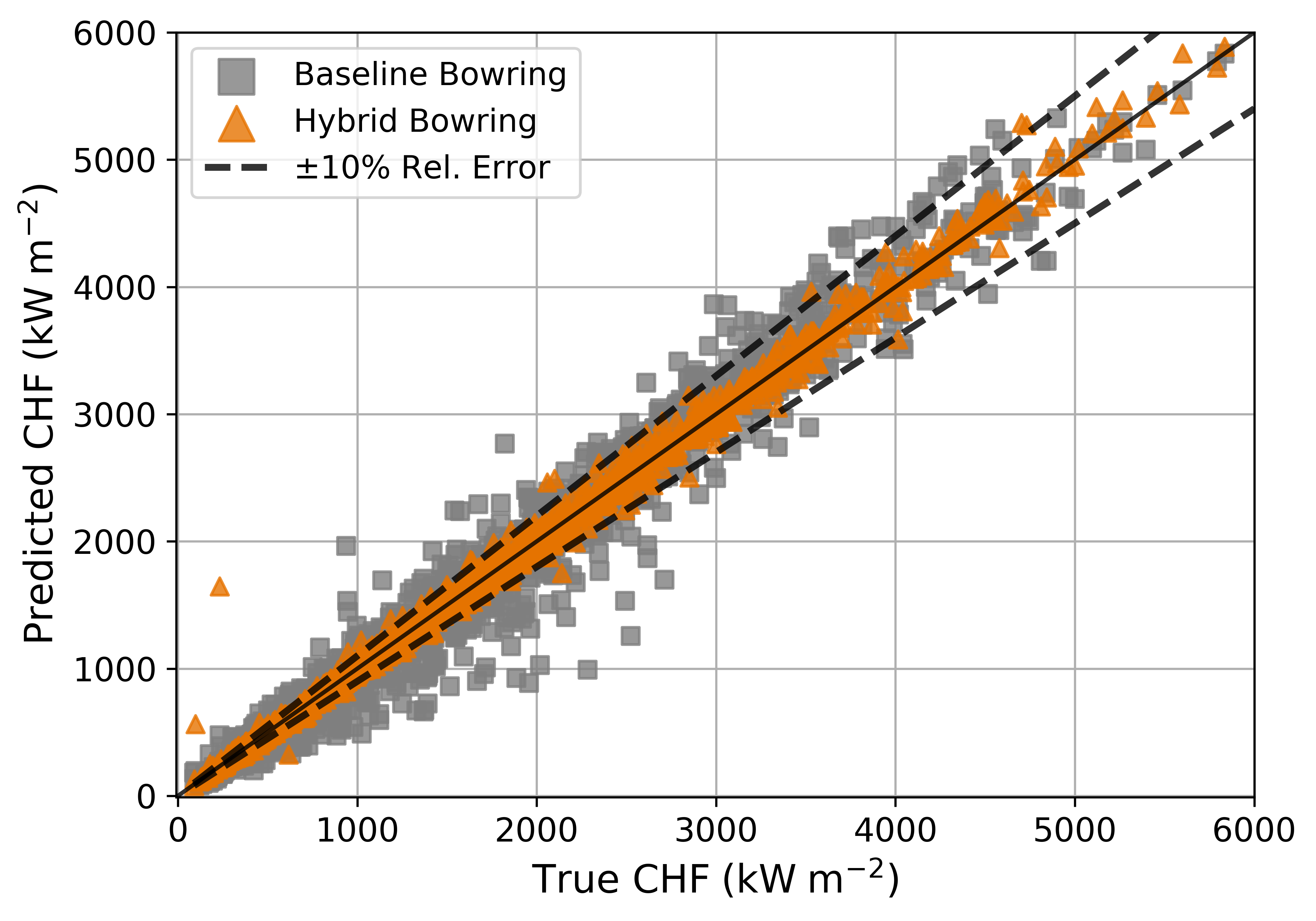}
        \caption{Parity}
    \end{subfigure}
    \begin{subfigure}{0.49\textwidth}
        \centering
        \includegraphics[width=0.96\linewidth]{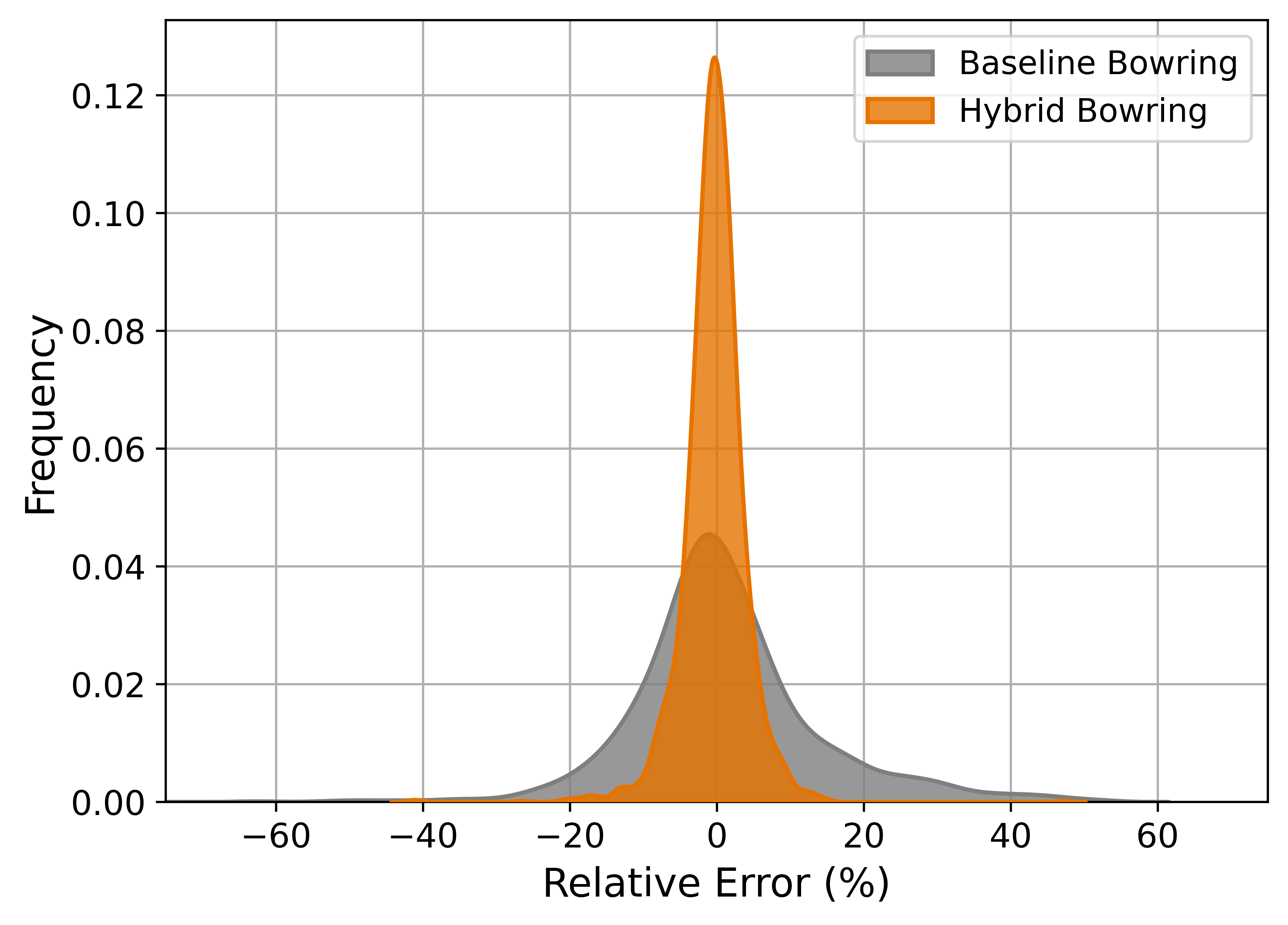}
        \caption{KDE plot of relative error}
    \end{subfigure}
    \caption{Comparisons between the baseline Bowring CHF model and its hybrid counterpart using the test partition of the NRC dataset. The parity plot is windowed at 6,000 \si{\kilo\watt\per\square\meter} for ease of interpretation, and the KDE plot is windowed at $\pm 70\%$.}
    \label{fig:nrc_bowring}
\end{figure}

To inspect how the hybrid ML models improve upon the baseline correlations, both the hybrid Biasi and hybrid Bowring results were plotted against their standalone counterparts, as shown in Figure \ref{fig:nrc_biasi} and Figure \ref{fig:nrc_bowring}. The hybrid Biasi predictions contain a similar region of larger error in the smaller CHF values in comparison with the baseline Biasi, but to a lesser degree. There appears to be active suppression of the overprediction bias of the Biasi correlation, with the exception of multiple extreme outliers at small CHF values. This general behavior is supported by the KDE plot, which shows a shift in the central tendency towards zero. The rest of the hybrid Biasi model's predictions in the parity plot are well-centered and are symmetric about the identity line along the \textit{x}-axis, maintaining a tighter spread in comparison with the standalone Biasi correlation's predictions. For the Bowring cases, there is an apparent decrease in error when comparing the hybrid model and the baseline Bowring correlation. With the exception of the two large outliers noted below 300 \si{\kilo\watt\per\square\meter}, the distribution of error is tight and avoids the underprediction of the Bowring correlation between 1,000 and 2,000 \si{\kilo\watt\per\square\meter}. The KDE plot shows the hybrid Bowring's relative errors centered almost exactly at zero, with a far smaller spread and better symmetry in comparison with the baseline Bowring predictions.

Compared with the baseline Biasi and Bowring correlations, all three ML-based models exhibited improved performance. The hybrid Bowring model exhibited the most favorable error metrics overall, followed by the pure ML model and then the hybrid Biasi model. The results from this NRC test indicate that these ML-based models could be useful augmentations to the current CHF models, with the hybrid approach offering improved interpretability in comparison with the purely data-driven model.

\subsection{Testing Results Using the Bennett Dryout Experiments}
\label{subsec:bennett_test}

The 224 CTF models of the Bennett test series were constructed and run in the same manner as the first test of Section \ref{subsec:nrc_test}, with the assumption that CHF occurred at the DO location. The DO location was identified by a marked increase in thermocouple temperature, consistent with the present CTF validation methodology for this test \cite{salko2023ctf}. This method of CHF identification is less precise in comparison with the NRC dataset, which provides the actual height at which the critical condition occurs. Due to the position of the Bennett experiments' thermocouples, this allows for a resolution of 3 to 12 in.~(0.076~\si{\meter} to 0.305 \si{\meter}), where CHF could have occurred anywhere between these thermocouple locations.

The same set of six error metrics was carried over from Section \ref{subsec:nrc_test}. The Bennett test's results are presented in Table \ref{tab:comparison_bennett}. There were multiple instances (18 of 224 cases) of zero-valued CHF predictions made by the baseline correlations, especially with the Biasi correlation. The cause is currently unknown, does not follow any apparent trends related to the input features, and is under further investigation. Because these large error values would lead to uninformative results from outlier-sensitive metrics, they were omitted when computing the mean relative error, the error's standard deviation, and the $rRMSE$. These values were \textit{not} omitted when computing the other metrics. The pure ML model's error values are larger than those of the NRC CHF database test, but all metrics are still more favorable than those of the standalone correlations. The baseline Biasi and Bowring cases saw mean relative errors of 9.49\% and 9.43\%, respectively, which are larger than the mean relative errors of any of the three ML-based models. The hybrid Biasi's predictions exhibited the best performance metrics of any model, with a mean relative error of 3.64\% and only 9.82\% of all points above 10\% error. The hybrid Bowring model's metrics fell in between those of the hybrid Biasi and pure ML's, with 21.43\% of points falling outside the 10\% boundary and a mean error of 7.88\%. No absolute error values above 25\% were observed in either of the hybrid ML models' predictions, with only 6.25\% in the pure ML's case. Although the pure ML model has smaller values than the standalone correlations in five of the six metrics, it has a similar number of points above 10\% error in comparison with the baseline Bowring.

\begin{table}[htb!]
    \centering
    \caption{Performance metrics for the Bennett experiments' CHF values. Zero-valued CHF points were removed when calculating $\upmu_\text{error}$, $\text{Std}_{\text{error}}$, and $rRMSE$.}
    \label{tab:comparison_bennett}
    \begin{tabular}{lccccc}
        \toprule
        Metric & Base & Base & Pure & Hybrid & Hybrid \\
         & Biasi & Bowring & ML & Biasi & Bowring \\
        \midrule
        $\upmu_\text{error}$ (\%) & $9.49$ & $9.43$ & $7.76$ & \cellcolor{gray!25}$3.64$ & $6.11$ \\ \hline
        $\text{Max}_{\text{error}}$ (\%) & $100.0$ & $100.0$ & $32.26$ & \cellcolor{gray!25}$18.33$ & $24.37$ \\ \hline
        $\text{Std}_{\text{error}}$ (\%) & $6.53$ & $12.41$ & $7.77$ & \cellcolor{gray!25}$3.67$ & $6.34$ \\ \hline
        $rRMSE$ (\%) & $11.51$ & $15.57$ & $10.97$ & \cellcolor{gray!25}$5.16$ & $8.79$ \\ \hline
        $F_{\text{error}}>10\%$ (\%) & $45.09$ & $30.80$ & $28.13$ & \cellcolor{gray!25}$9.82$ & $21.43$ \\ \hline
        $F_{\text{error}}>25\%$ (\%) & $9.38$ & $13.84$ & $6.25$ & \cellcolor{gray!25}$0.00$ & \cellcolor{gray!25}$0.00$ \\
        \bottomrule
    \end{tabular}
\end{table}

\begin{figure}[htb!]
    \centering
    \begin{subfigure}{0.49\textwidth}
        \centering
        \includegraphics[width=\linewidth]{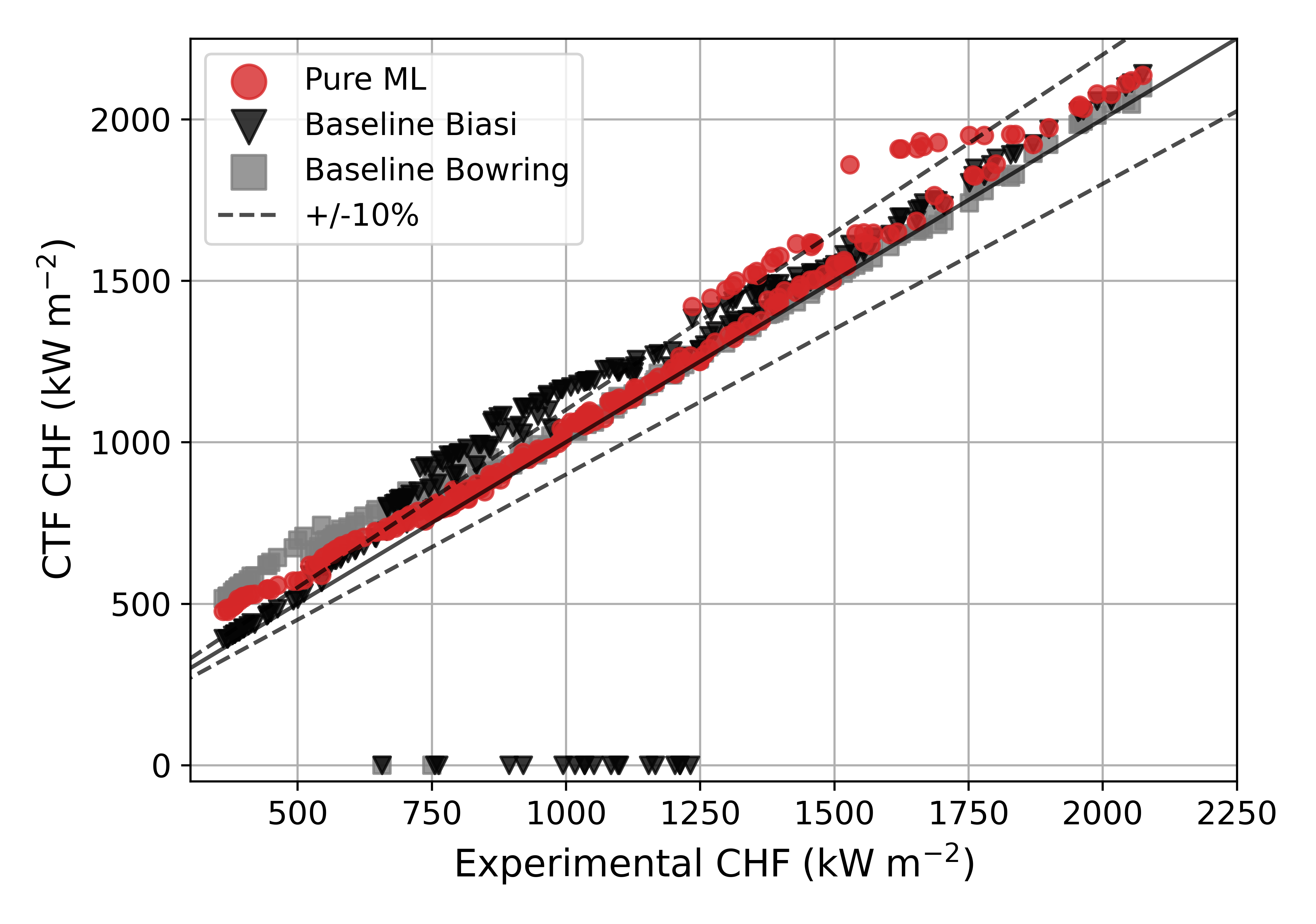}
        \caption{Parity}
        \label{subfig:bennett_parity_pure}
    \end{subfigure}
    \begin{subfigure}{0.49\textwidth}
        \centering
        \includegraphics[width=0.967\linewidth]{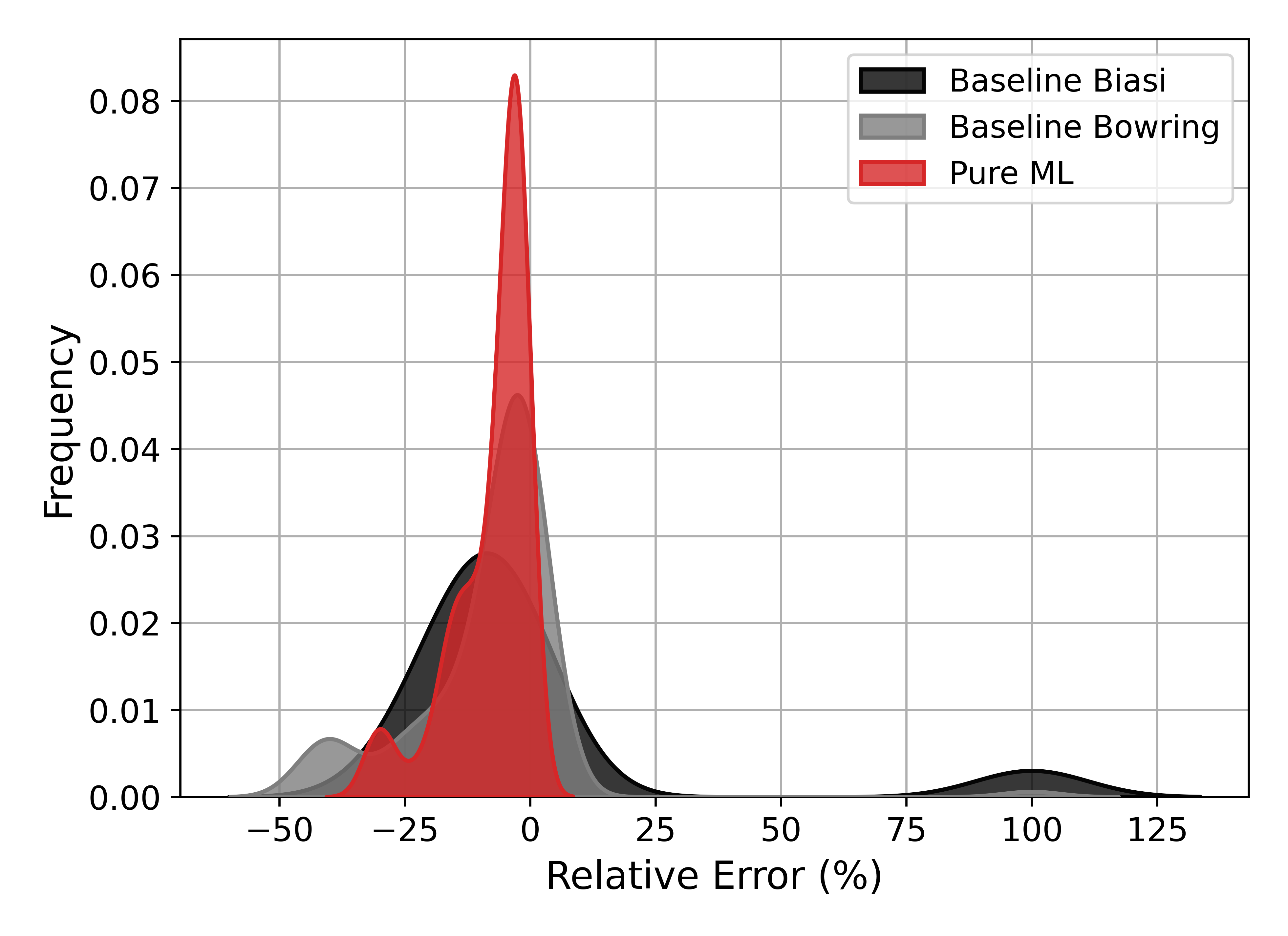}
        \caption{KDE plot of relative error}
        \label{subfig:bennett_kde_pure}
    \end{subfigure}
    \caption{Parity and KDE comparisons between baseline CHF models and the pure ML predictions using the Bennett test series.}
    \label{fig:bennett_parity_kde_pure}
\end{figure}

To provide a visualization of the agreement of these predictions with experimental values, parity plots were once again used. The first comparison involves the standalone Biasi and Bowring correlations and the pure ML model in Figure \ref{fig:bennett_parity_kde_pure}. Both the baseline cases are observed with deviation in the lower CHF value region, consistently overpredicting values. Most of the disagreement for the Biasi correlation is between 750 and 1,250 \si{\kilo\watt\per\square\meter}, and most of the disagreement for the Bowring correlation is between 300 and 850 \si{\kilo\watt\per\square\meter}. The Biasi correlation also predicted several negative CHF values, which are corrected to zero within CTF to remain somewhat physically consistent for the simulation. The standalone Bowring correlation also saw two points at zero. The pure ML model has a more inconsistent pattern of disagreement but maintains stronger accuracy overall. Each of the three scatters indicates slight overprediction, even on the segments close to the identity line. This could be due to how the experimental CHF values were extracted because the critical condition would likely occur between two thermocouples. The KDE plot (Figure \ref{subfig:bennett_kde_pure}) shows the distribution of relative error values. The pure ML model has a tighter distribution in comparison with the baseline models, albeit with a small overprediction bias in comparison with Bowring. The outlier values of the standalone correlations are located on the far right, represented by the small symmetric distribution centered at 100\%. The KDE plot shows a small density extension beyond 100\% due to kernel smoothing. Because the highest observed errors were clustered just below 100\%, the Gaussian kernel distributes some probability mass slightly beyond this range.

\begin{figure}[htb!]
    \centering
    \begin{subfigure}{0.49\textwidth}
        \centering
        \includegraphics[width=\linewidth]{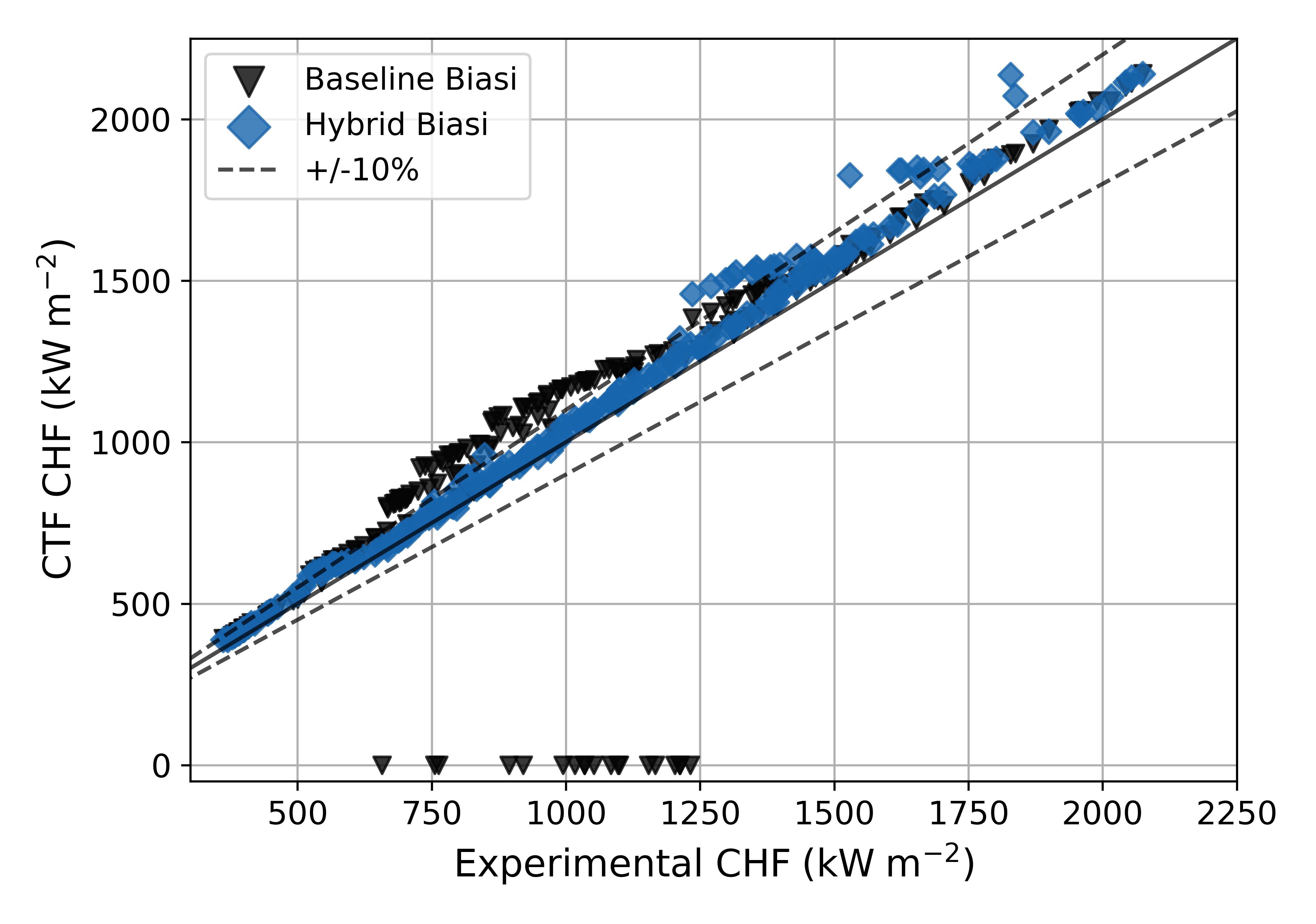}
        \caption{Parity}
        \label{subfig:bennett_parity_biasi}
    \end{subfigure}
    \begin{subfigure}{0.49\textwidth}
        \centering
        \includegraphics[width=0.967\linewidth]{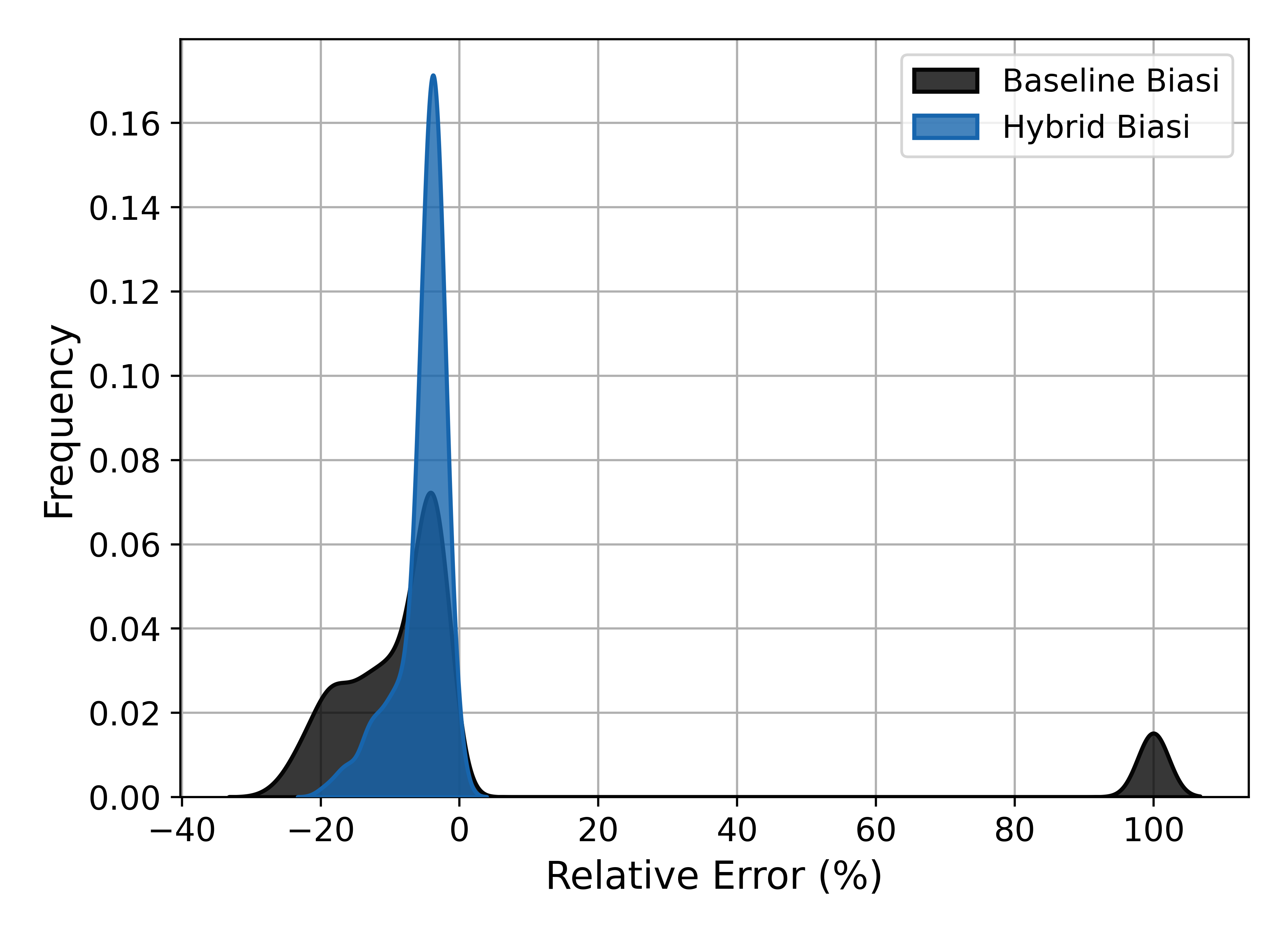}
        \caption{KDE plot of relative error}
        \label{subfig:bennett_kde_biasi}
    \end{subfigure}
    \caption{Parity and KDE comparisons between the baseline and hybrid Biasi models using the Bennett test series.}
    \label{fig:bennett_parity_kde_biasi}
\end{figure}

\begin{figure}[htb!]
    \centering
    \begin{subfigure}{0.49\textwidth}
        \centering
        \includegraphics[width=\linewidth]{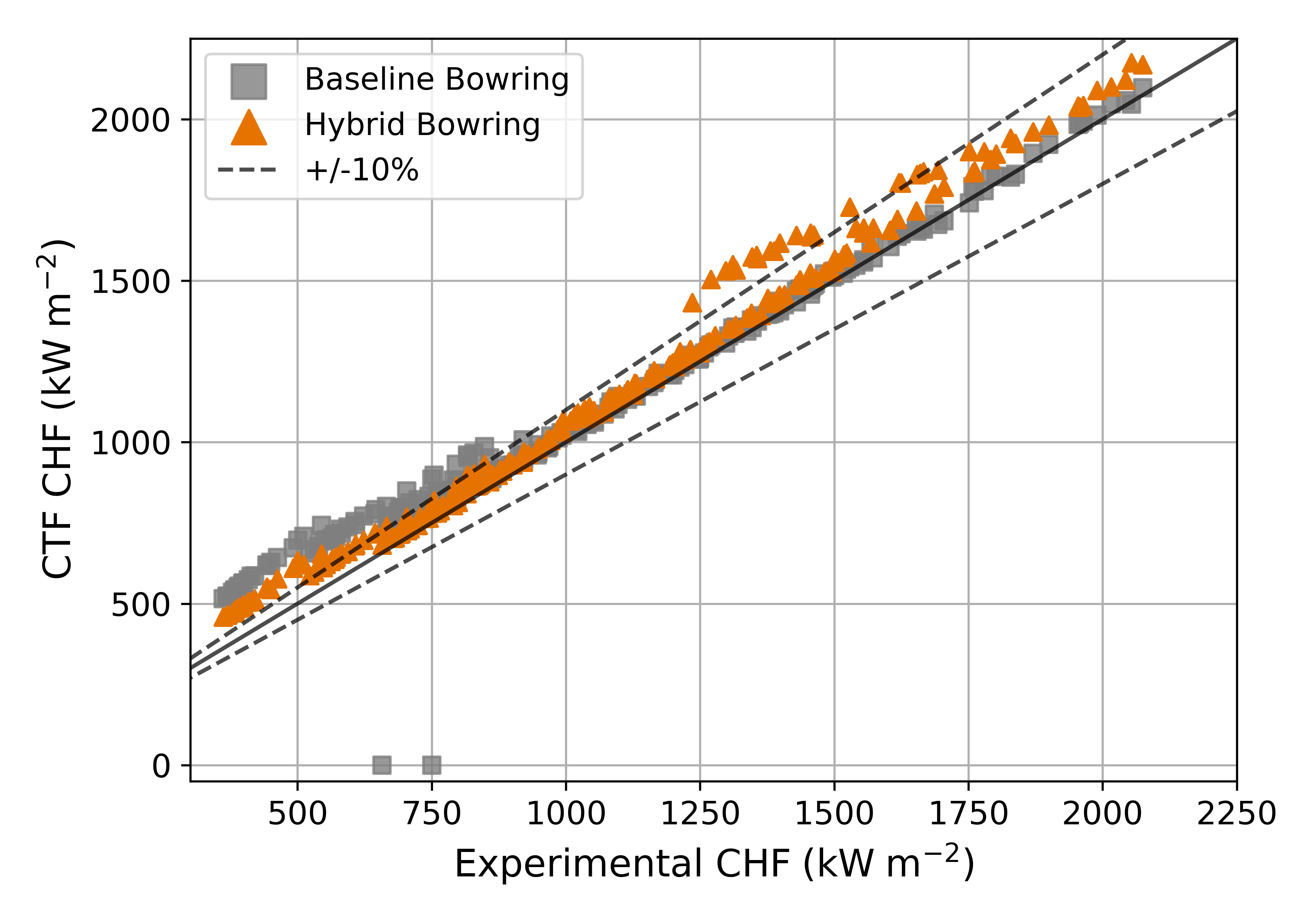}
        \caption{Parity}
        \label{subfig:bennett_parity_bowring}
    \end{subfigure}
    \begin{subfigure}{0.49\textwidth}
        \centering
        \includegraphics[width=0.983\linewidth]{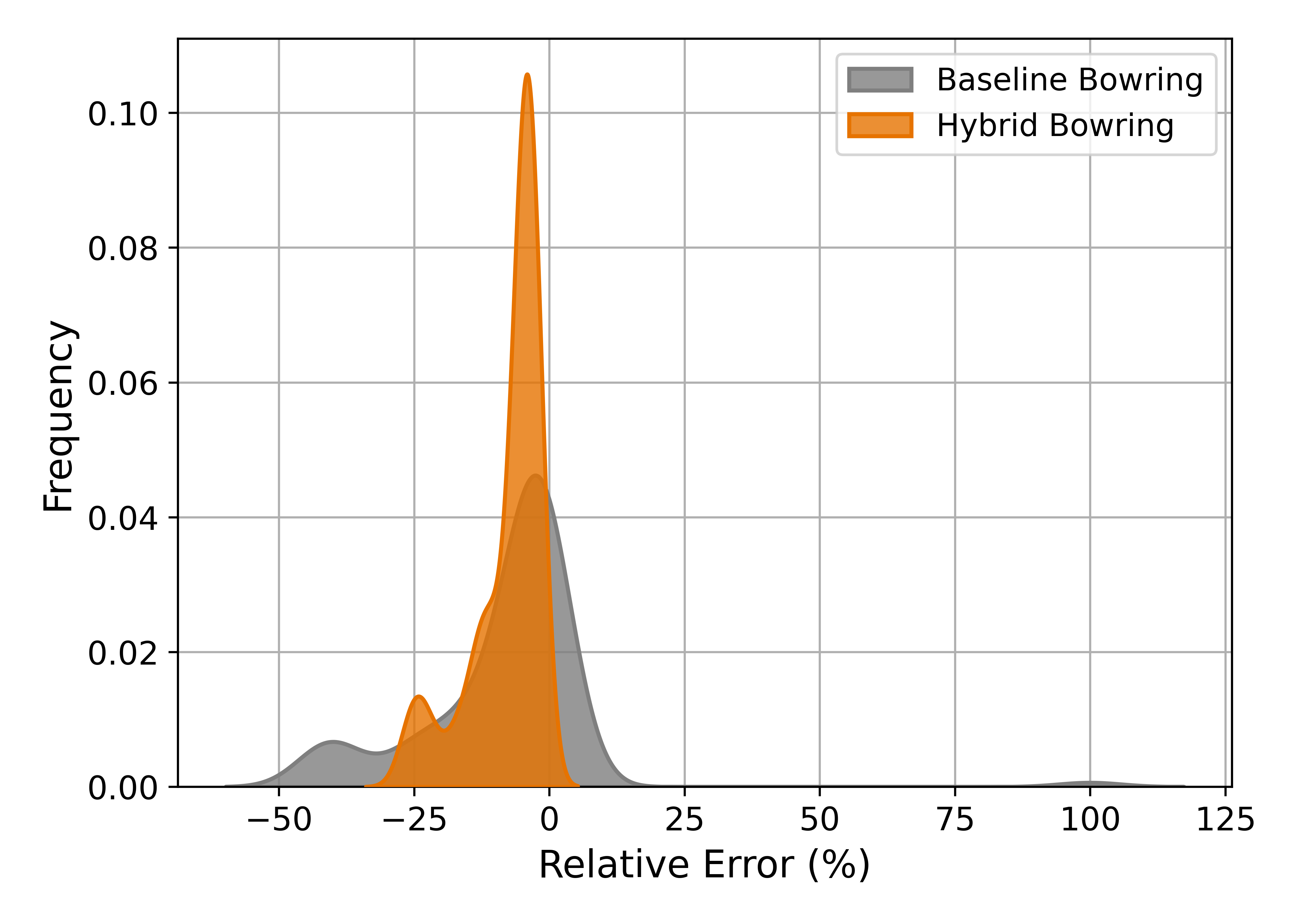}
        \caption{KDE plot of relative error}
        \label{subfig:bennett_kde_bowring}
    \end{subfigure}
    \caption{Parity and KDE comparisons between the baseline and hybrid Bowring models using the Bennett test series.}
    \label{fig:bennett_parity_kde_bowring}
\end{figure}

Each of the hybrid models was then compared against its standalone base model, as shown in Figure \ref{fig:bennett_parity_kde_biasi} and Figure \ref{fig:bennett_parity_kde_bowring}. For the Biasi cases, the hybrid model successfully suppressed the overprediction of the Biasi correlation in the center experimental values, showing good agreement with the identity line. A limited number of hybrid-predicted points remain above the 10\% boundary at higher \textit{x}-axis values and a small segment between 1,250 and 1,500 \si{\kilo\watt\per\square\meter}. Generally, there is significant improvement in adherence to the identity line when using the hybrid ML model in comparison with the baseline Biasi correlation. This is supported by the KDE plot, which shows the Biasi error distribution tightly centered just below zero, with some left-hand mass representing the high-CHF overprediction. The baseline Biasi distribution is shallow and wide, with a heavier bias for overprediction. For the Bowring cases, the departure from the identity line by the standalone Bowring correlation is notably reduced with the hybrid method. At relatively small CHF values (below 600 \si{\kilo\watt\per\square\meter}), the hybrid model has a tendency to overpredict values but not to the same extent as the baseline correlation. Along most of the \textit{x}-range, the hybrid model shows strong agreement with the ideal state, albeit with some overprediction between 1,250 and 1,500 \si{\kilo\watt\per\square\meter}. The distribution of relative error for the hybrid case is tighter than that of the standalone Bowring, as evidenced by the KDE plot. Both are centered at nearly the same position and are slightly biased for overprediction, but the standalone Bowring has a larger spread on both sides of the hybrid distribution.

\section{Conclusions}
\label{sec:conclusions}

This study investigated the integration of ML techniques into the CTF subchannel code for CHF prediction. Three ML-based models were implemented via a native Fortran framework: a purely data-driven DNN and two hybrid approaches that combined empirical correlations (Biasi and Bowring) with ML-based residual correction. The use of hybrid approaches provides physics knowledge to a model, reducing the amount of inferred knowledge required by the ML component and thereby increasing interpretabilty and resistance to data scarcity. All three models were trained using a partition of the public NRC CHF database originally used to construct the 2006 Groeneveld LUT. These models were then evaluated against experimental CHF data from a subset of the NRC database and the Bennett DO experiments.

Results for a set of six error metrics indicate that all the ML-based models offer improved performance in comparison with standalone empirical correlations. In both testing cases, the hybrid models mitigated the discrepancies present in their standalone correlation counterparts. The pure ML model remained competitive with the hybrid models, consistently exhibiting reduced error in comparison with the empirical correlations in nearly all metrics. Parity and KDE plots indicate that traditional correlations (particularly Biasi) tend to overpredict CHF in lower-value regions, whereas ML models mitigate this bias. The hybrid models were shown to improve generalization by correcting correlation deficiencies without fully discarding physical bases.

Future work will explore the CTF implementation of uncertainty quantification methods, such as Bayesian neural networks and deep ensembles, to provide built-in predictive uncertainty estimates. Additional investigations will focus on extending the hybrid approach to other models and applying it to broader benchmark datasets, such as the BWR Full-size Fine-mesh Bundle Test (BFBT) and PWR Sub-channel Bundle Test (PSBT). The use of transfer learning is also under active investigation to leverage the NRC database trained models (tube geometry) for use in alternative geometries, such as annuli \cite{BECKER198147} or rectangular channels \cite{furlong2024transfer}. Overall, this study's findings underscore the potential of ML-based models as augmentations to traditional CHF prediction methods. By improving predictive accuracy while maintaining interpretability, hybrid ML approaches present a viable path toward more reliable and adaptable CHF modeling frameworks in nuclear reactor simulations.

\section*{Acknowledgments}

This research was supported in part by an appointment to the US Department of Energy's (DOE's) Omni Technology Alliance Internship Program, sponsored by DOE and administered by the Oak Ridge Institute for Science and Education (ORISE). This work was also partially funded by DOE's Office of Nuclear Energy Distinguished Early Career Program (DECP) under award number DE-NE0009467.

\newpage
\bibliography{./bibliography.bib}

\end{document}